\documentclass[letterpaper,aps,prl,reprint,superscriptaddress,showpacs]{revtex4-1}

\usepackage[pass]{geometry}
\usepackage{amsfonts,amsmath,amssymb,color,graphicx} % assumes amsmath package installed
\usepackage{subfigure}

\usepackage{graphicx}% Include figure files
\usepackage{epstopdf}
\DeclareGraphicsExtensions{.eps,.png}

\newcommand{\dif}{\,\mathrm{d}}		  % Define dif
\newcommand{\der}[2]{\mathop{\frac{d #1}{d #2}}}
\newcommand{\midop}[1]{\mathop{\langle #1 \rangle}\nolimits}
\renewcommand{\section}[1]{\emph{#1.}}
\newcommand{\xForce}{G}
\newcommand{\pForce}{F}
\newcommand{\Nbath}{K}
\usepackage{color}
\usepackage[active,generate=equations,extract-env={equation,align}]{extract}

\begin{document}

\title{Non-conservative Forces via Quantum Reservoir Engineering}

\author{Shanon L. Vuglar}
\affiliation{University of Melbourne, Parkville, VIC 3010, Australia }
\affiliation{Princeton University, Princeton, NJ 08544, USA}

\author{Dmitry V. Zhdanov}
\affiliation{Northwestern University, Evanston, IL 60208, USA}

\author{Renan Cabrera}
\affiliation{Princeton University, Princeton, NJ 08544, USA}

\author{Tamar Seideman}
\affiliation{Northwestern University, Evanston, IL 60208, USA}

\author{Christopher Jarzynski}
\affiliation{University of Maryland, College Park, MD, 20742, USA}

%\author{Herschel A. Rabitz}
%\affiliation{Princeton University, Princeton, NJ 08544, USA}

\author{Denys I. Bondar}
\affiliation{Princeton University, Princeton, NJ 08544, USA}

\date{\today}

%%%%%%%%%^%%%%%%%%%^%%%%%%%%%^%%%%%%%%%^%%%%%%%%%^%%%%%%%%%^%%%%%%%%%^%%%%%%%%%^
\begin{abstract}

A systematic approach is given for engineering dissipative environments that steer quantum wavepackets along desired trajectories. The methodology is demonstrated with several illustrative examples: environment-assisted tunneling, trapping, effective mass assignment and pseudo-relativistic behavior. Non-conservative stochastic forces do not inevitably lead to decoherence -- we show that purity can be well-preserved.
These findings highlight the flexibility offered by non-equilibrium open 
quantum dynamics.

%As in classical statistical physics, the effect of a dissipative environment on a quantum probe can be described by a quantum statistical force. Using Operational Dynamic Modeling, we show how to designe an environment to exert a desired quantum statistical force on a quantum probe. As examples, we simulate environments tailored 
%to enhance quantum tunneling,
%to trap particles, 
%to induce effective mass, and 
%to produce quasi-relativistic dynamics. 
%We also provide a possible experimental implementation for this environment. 
%These findings highlight the flexibility offered by non-equilibrium open 
%quantum dynamics.
%\vspace{1.cm}
\end{abstract}

%\keywords{}

\maketitle

%\subsubsection{Introduction}

%%%%%%%%%%%%%%%%%%%%%%%%%%%%%%%%%%%%%%%%%%%%%%%%%%%%%%%%%%%%%%%%%%%%%%%%%%%%%%%%%%%%%%%%%%%%%%%

\section{Introduction} 
Throughout its short history, the control of quantum systems has predominantly been implemented using conservative forces, e.g., manipulating quantum phenomena in Hamiltonian systems via dipole coupling with laser or microwave pulses.
This may seem surprising given the widespread use of non-conservative forces in other control applications -- consider the wind (sailing vessels, windmills) and friction (mechanical brakes).
The historical focus on conservative forces is, perhaps, best explained
by the widely held belief that immersing a quantum system into a complex environment inevitably destroys its quantum dynamical features. 
The monopoly of conservative forces in quantum control is now being challenged by quantum reservoir engineering (QRE)
\cite{poyatos1996quantum,verstraete2009quantum,fedortchenko2014finite,kurizki2015thermal,kienzler2015quantum,pan2016ground,rouchon2014models}. In particular, it has been shown that it is possible to preserve and even enhance the quantum dynamical features of a system by judiciously coupling the system to a dissipative environment.
Applications of quantum reservoir engineering 
include amplification \cite{metelmann2014quantum},
nonreciprocal photon transmission \cite{metelmann2015nonreciprocal,metelmann2017nonreciprocal},
photon blockade \cite{miranowicz2014state},
efficient photoinduced charge separation in solar energy conversion \cite{Zhdanov2015quantum},
binding of atoms \cite{lemeshko2013dissipative,wuster2017quantum},
inducing phase transitions \cite{kaczmarczyk2016dissipative,weimer2016tailored,overbeck2017multicritical}, 
implementation of quantum gates \cite{albert2016holonomic,ticozzi2017quantum,arenz2016universal,arenz2017lindbladian},
and the generation of entangled 
\cite{cheng2016preservation,liu2016comparing,zippilli2015steady,yang2015generation,mirza2015controlling,arenz2013generation},
squeezed \cite{kronwald2013arbitrarily,woolley2014two,grimsmo2016quantum}, 
and other exotic 
\cite{koga2012dissipation,holland2015single,asjad2014reservoir,chestnov2016permanent} quantum states.

In this Letter, we provide a systematic approach for engineering dissipative environments that steer quantum wavepackets along desired trajectories as defined by the following equations: 
\begin{subequations}\label{__problem_gen}
\begin{gather}
\der{}{t}{\midop{\hat x}}{=}\midop{\xForce(\hat p)}, \label{eqn:Ehrenfest1}\\
\der{}{t}{\midop{\hat p}}{=}\midop{\pForce(\hat x)}.  \label{eqn:Ehrenfest2}
\end{gather}
\end{subequations}
Here, $\midop{\hat x}$ and $\midop{\hat p}$ denote the wavepacket's mean position and momentum.
The environments obtained not only enhance desired quantum properties, 
but can also be made to preserve the purity of the underlying quantum system.
Equations~\eqref{__problem_gen} with various functions $\xForce$ and $\pForce$ embrace a plethora of quantum behaviors; we provide several illustrative examples. 
We first consider compensating for a potential barrier in the case of quantum tunneling and then mimicking a potential to trap a wave packet at a desired location. 
We also consider more exotic applications such as changing the effective mass of a quantum particle and emulating relativistic effects. % at low velocities $\der{}{t}\midop{\hat x}{\ll}c$. 
The scope of our analysis is restricted to Markovian environments modeled within the Lindblad formalism. 
We also discuss possible laboratory realizations of the Lindblad operators for specific examples. 

%%%%%%%%%%%%%%%%%%%%%%%%%%%%%%%%%%%%%%%%%%%%%%%%%%%%%%%%%%%%%%%%%%%%%%%%%%%%%%%%%%%%%%%%%%%%%%%

\section{Formal analysis}
For definiteness, assume that the system of interest is a one-dimensional particle of mass $m$ moving in a potential $U(x)$. Our objective is to dissipatively couple the system to $K+N$ baths in such a way that the average particle localization in phase space will follow Eqs.~\eqref{__problem_gen} for given, desirable, $\xForce(\hat p)$ and $\pForce(\hat x)$. Assuming %memoryless 
Markovian system-bath interactions, the system state (described by the density matrix $\hat\rho$) evolves according to the Lindblad master equation
\begin{align}
\frac{\dif \hat{\rho}}{\dif t} &= 
    - \frac{i}{\hbar} [ \hat{H}, \hat{\rho} ] 
	+ \sum_{k{=}1}^{\Nbath}\mathcal{D}_{\hat{A}_k}[\hat{\rho}]
    + \sum_{n{=}1}^{N}\mathcal{D}_{\hat{B}_n}[\hat{\rho}]
	, \label{eqn:dW}
\end{align}
where $\hat H$ is a given system Hamiltonian
\begin{align}
\hat{H} &= \frac{1}{2m}{\hat{p}}^2 + U(\hat{x}), \label{eqn:H}
\end{align}
and the effect of the bath is represented via the operators 
$\hat A_k$, $\hat B_n$ as
\begin{align}
\mathcal{D}_{\hat{A}}[\hat{\rho}] &= 
	\frac{1}{\hbar} \Big( \hat{A} \hat{\rho} \hat{A}^\dagger 
		 -	\frac{1}{2} \hat{\rho} \hat{A}^\dagger \hat{A}  
			-	\frac{1}{2} \hat{A}^\dagger \hat{A} \hat{\rho} 
			\Big). \label{eqn:D}
\end{align}
Under these assumptions, the control problem reduces to determining suitable forms for 
the operators $\hat A_k$, $\hat B_n$ 
and providing physical evidence that the corresponding environments can be engineered 
in the laboratory. 
Using Operational Dynamical Modeling \cite{Bondar2011c,Bondar2014wigner} the following expressions for $\hat{A}_k = A_k(\hat{x})$ and $\hat{B}_n = B_n(\hat{p})$ are obtained:
\newcommand{\BWV}{{\cal K}}% Bondar wavevector
\begin{subequations}\label{eqn:sol}
\begin{align}
A_k(x) = \: &R_k(x) \exp \left( 
                {i\int \frac{f_k(x)}{R_k^2(x)} \dif x} 
            \right), \label{eqn:A} \\
B_n(p) = \: &S_n(p) \exp \left( 
                {-i \int \frac{g_n(p)}{S_n^2(p)} \dif p}
            \right). \label{eqn:B}
\end{align}
Here, $f_k(x)$, $g_k(p)$, $R_k(x)$, and $S_k(p)$ 
denote arbitrary real valued functions such that 
\begin{gather}\label{eqn:F_k}
\sum_{k=1}^K f_k(x){=}\pForce(x){+}\der{U(x)}{x};~~\sum_{n{=}1}^{N} g_n(p){=}\xForce(p){-}\frac{p}{m}.
\end{gather}
\end{subequations}
%
%The physical nature of environments that implement the solutions \eqref{eqn:sol} varies substantially with specific choices of the operators $\xForce$ and $\pForce$. 
%
Note that Eqs.~\eqref{__problem_gen} are satisfied regardless of the initial state.
% TODO rephrase below
To provide insight into the physical nature of environments that implement (5), we now consider several illustrative examples. 
Unless stated otherwise, atomic units (a.u.),  $\hbar = m_e = |e| = 1$, are used throughout.

%%%%%%%%%%%%%%%%%%%%%%%%%%%%%%%%%%%%%%%%%%%%%%%%%%%%%%%%%%%%%%%%%%%%%%%%%%%%%%%%%%%%%%%%%%%%%%%

\section{Environmentally assisted quantum tunneling}
\begin{figure}[tbp]
\centering\includegraphics[width=0.7\columnwidth]
{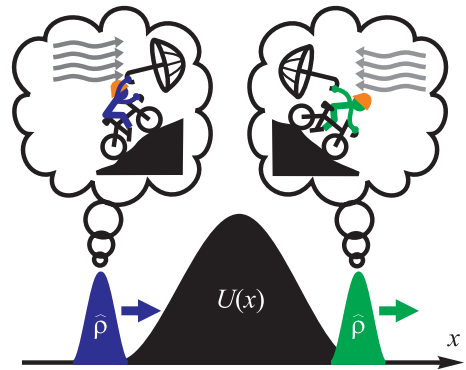}
\caption{Environment assisted quantum tunneling resembles cycling with an umbrella: The environment action is qualitatively similar to tailwind (headwind) when going uphill (downhill). The net effect is a reduction of the back-scattering probability with minimal side effects on the wavepacket parameters.
\label{@FIG.01}}
\end{figure}

It is common knowledge that cycling uphill is much easier with assistance from a tailwind. 
Similarly, a ``polarized electron wind'' can be used to enhance tunneling rates for an atomic wavepacket approaching a potential barrier $U(\hat x)$ (see Fig.~\ref{@FIG.01}).
If non-conservative forces are engineered so as to cancel the potential forces of the system, then dynamics similar to those of a free particle can be obtained.
Consider Eqs.~\eqref{__problem_gen} and choose $\xForce(p){=}\frac{p}{m}$ and $\pForce(\hat x){=}0$. 
These dynamics can be obtained with the following choice of environmental operators $A_{\pm}$, which satisfy \eqref{eqn:sol} for the case 
$K=2$, $N=0$, and $R_1 = R_2 = C$ where $C$ is a constant:
\begin{subequations}\label{eqn:sol_tun}
\begin{gather}
A_{\pm}{=}C e^{\pm\frac{2i}{\hbar}\int{{\tilde p_{\pm}(x)d x}}}, \label{eqn:A_tun}
\end{gather}
where the functions $\tilde p_{\pm}(x)$ obey the relation
\begin{gather}
\tilde p_{+}(x){-}\tilde p_{-}(x){=}\tfrac{\hbar}{2C^2}\tfrac{\dif U(x)} {\dif x}. \label{eqn:p_pm_tun}
\end{gather}
\end{subequations}
%Note that while dissipative forces are typically thought of as ones which convert a system's energy into heat, this is not the case here.

Inspired by the wind analogy, we now propose a physical implementation of the environment~\eqref{eqn:sol_tun}. 
Consider a quantum probe that is an atom of mass $m$ in the non-degenerate ground electronic state with electric polarizability $\alpha$, zero angular momentum, and negligible magnetic polarizability. 
Suppose that the motion of the probe along the $\vec\epsilon_x$-axis is impeded by an effective barrier $U(x)=-\alpha {\cal E}(x)^2/4$ 
created by an off-resonant, blue-detuned ($\alpha{<}0$) laser field ${\vec\epsilon}_x \mathcal{E}(x)\cos(\omega(t-{z}/{c}))$. 
In the presence of a static magnetic field of the form ${\vec\epsilon}_z{\cal B}(x)$, the desired dissipative environment can be created by two counterpropagating electron jets, in which the electrons have opposite magnetic moments 
$\hat{\mu }_s{=}{\pm}\hat{\sigma }_z  \mu_{\mathrm{B}}$, incident velocities ${\pm}{\vec\epsilon }_x\frac{p_0}{m_{\mathrm{e}}}$, and fluxes ${\pm}{\vec\epsilon }_x j$ (here $\mu _{\mathrm{B}}$ is the Bohr magneton).
The resulting electron recoils create an effective pressure on the probe. 
Note that without a magnetic field, the mean impacts of both jets would mutually compensate each other. 
However, when a magnetic field is applied, the opposite electron spin polarizations of the jets break this symmetry resulting in a nonzero net force on the probe. 

To quantitatively describe this effect, we assume that~i) the electron flux $j$ is low enough to neglect multiple scattering of electrons,~ii) all interactions of electrons with the probe can be modeled as ideal elastic backscattering events,~iii) the incident electron velocity ${p_0}/{m_{\mathrm{e}}}$ is much larger than the characteristic velocities of the probe, and~iv) 
\begin{equation}
    p_0{\gg}\sqrt{2\mu_{\mathrm{B}}m_{\mathrm{e}}|{\cal B}(x)|} \label{eqn:pgg}.
\end{equation}
The inequality \eqref{eqn:pgg} allows the wavefunctions of incident electrons in the jets to 
be modeled semiclassically as
\begin{equation}
\psi_{\pm}{\propto}\frac{e^{\frac{\pm i}{\hbar}\int \tilde p_{\pm}(x)\dif x}}{\tilde p_{\pm}(x)},
~~\tilde p_{\pm}(x){=}\sqrt{p_0^2\pm 2\mu_{\mathrm{B}}m_{\mathrm{e}}{\cal B}(x)}. \label{eqn:ptilde}
\end{equation}
In the case of $C = \sqrt{\hbar \tilde{\sigma} j}$ where $\tilde\sigma$ is the scattering cross section, Eqs. (\ref{eqn:dW}) and \eqref{eqn:A_tun} 
describe the ``wind effect'' of the electron jets on the probe. 
Note that $C^2$ is proportional to the number of electron scatterings in a given time interval. Under the assumption of Poissonian statistics, the standard deviation over the same time interval of the force exerted by the collisions is expected to be proportional to $Cp_0$.
%
%The average incident energy of each electron jet on the probe per unit time is given by  
%$\tilde{\sigma} j p_0^2 / (2m_e) = (Cp_0)^2 / (2m_e \hbar)$.
%Hence, under the assumption of Poissonian collisional dynamics, $Cp_0$ is proportional to the standard deviation of the incident energy fluctuations. 
%
This parameter will be used below to elucidate physical mechanisms.
Finally, Eqs.~\eqref{eqn:p_pm_tun} and \eqref{eqn:ptilde} determine the magnetic field profile required for effectively barrierless propagation:
\begin{align}
{\cal B}(x) &=  \frac{\hbar}{16 \mu_\mathrm{B}m_\mathrm{e}C^4}
                {\frac{\dif U(x)}{\dif x} 
                \sqrt{16 {p_0}^2 C^4 - \left( \hbar \frac{\dif U(x)}{\dif x}\right)^2}}.
                \label{eqn:calB}
\end{align}

\begin{figure}
		\includegraphics[width=1\hsize]{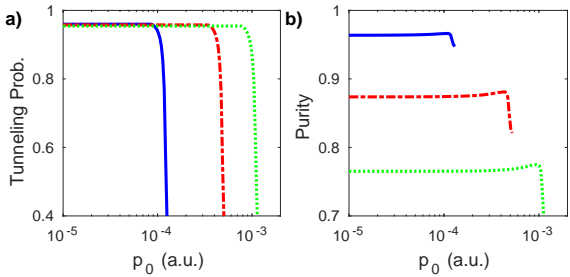}
        \caption{The transmission probability (a) and purity (b)  for a Gaussian atomic wavepacket tunneling through a potential barrier in the presence of electron jets [Eqs.~\eqref{eqn:dW},\eqref{eqn:sol_tun},\eqref{eqn:ptilde} and \eqref{eqn:calB}] as a function of electron momentum $p_0$. 
        In each case the wavepacket has initial mean kinetic energy $K_0 = 0.0068$ (a.u.) and the potential barrier is $U(x) =  2 K_0 e^{-x^2/2}$. 
        The solid blue, dash-dotted red, and dotted green curves correspond to incident electron energy fluctuations $C p_0 = 5 \times 10^{-4}$, $10^{-3}$ and $1.5 \times 10^{-3}$ (a.u.) respectively. 
        The right-most points of the curves in panel (b) correspond to the case when inequality \eqref{eqn:pgg} turns into equality.
        \label{fig:line_plots}}
\end{figure}

The character of the system-environment coupling is determined by the momenta $p_{\pm}$ of the incident electrons.
For small magnitudes of $|p_{\pm}|$, large collision rates are required to create sufficient non-conservative forces to oppose the potential forces. 
In this case, the overall effect of the collisions can be represented as an effective pressure, and the dissipative term in \eqref{eqn:dW} in the limit $C{\to}\infty$, $|p_{\pm}|{\to}0$ can be represented as an effective Hamiltonian,
$\hat{H}_{\mathrm{eff}} = -U(\hat{x})$,
which cancels the potential barrier $U(x)$ and results in entirely coherent (essentially free-particle) dynamics. 
On the other hand, large $|p_{\pm}|$ corresponds to the shot noise limit where strong but rare collisions produce highly fluctuating stochastic environmental forces. This leads to rapid wavepacket decoherence and a reduction in tunneling probabilities. 
These effects can be seen in Fig.~\ref{fig:line_plots}, 
which depicts simulation results for a hydrogen-like atom ($m{=}1837m_e$ where $m_e$ is electron mass) tunneling through a Gaussian potential barrier in the presence of an engineered environment as described by Eqs. (\ref{eqn:dW}), \eqref{eqn:sol_tun}, and \eqref{eqn:ptilde}. 
In all cases, Eqs.~\eqref{__problem_gen} are satisfied.
For small values of $|p_{\pm}|$, high tunneling rates and purity are achieved 
for the atomic quantum state after interaction with the barrier.  
However, above a critical $|p_{\pm}|$ 
(which depends on the standard deviation of the environmental force $Cp_0$), 
the tunneling probability and purity dramatically degrade; 
this corresponds to the shot noise regime.

One can observe slight increases in the purity prior to the rapid falling away in each of the curves in Fig.~\ref{fig:line_plots}(b). These peaks correspond to a transitional regime wherein the tunneling rates are starting to degrade [see Fig.~\ref{fig:line_plots}(a)] and reflection from the barrier becomes noticeable ($\propto 10\%$). 
Furthermore, the inequality \eqref{eqn:pgg} is only marginally satisfied; the observed purity increase may be an artifact of the semi-classical approximation \eqref{eqn:ptilde}.
%The emergence of interference fringes between the transmitted and reflected wavepackets results in an increase in purity.  

The ability of environmental coupling to enhance tunneling rates has been previously recognized. 
Under certain physical conditions, a metastable quantum system submerged into a low temperature environment decays, exciting directional bath modes such that the quantum system acquires kinetic energy which in turn assists under the barrier motion~\cite{leggett1984quantum,grabert1984quantum, pollak1986transition,leggett1996effect}. 
In particular, an atom can acquire an extra momentum kick, facilitating tunneling by spontaneously emitting a photon. 
This mechanism has been systematically explored in Refs. \cite{japha1996, Schaufler1999keyhole} and yielded Zeno and anti-Zeno quantum control schemes \cite{barone2004}. 
In these schemes the incident wavepackets undergo destructive spontaneous dissipative changes. However, in our example the enhanced tunneling is achieved without destroying the state purity, as can be seen from Fig.~\ref{fig:line_plots}.
It is noteworthy that environmentally assisted tunneling was recently 
experimentally demonstrated in 
lithium niobate \cite{somma2014high}. 

%%%%%%%%%%%%%%%%%%%%%%%%%%%%%%%%%%%%%%%%%%%%%%%%%%%%%%%%%%%%%%%%%%%%%%%%%%

\section{Dissipative traps} 
\begin{figure}
		\includegraphics[width=1\hsize]{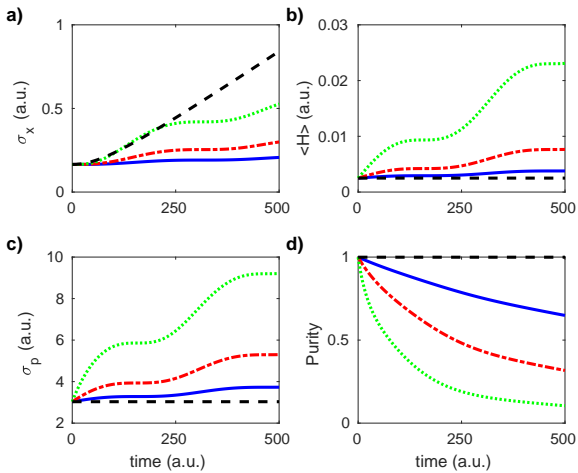}
        \caption{The spreading in position (a), 
        increase in energy (b), 
        spreading in momentum (c), and 
        decrease in purity (d) 
        for an atomic wavepacket initially in the ground state of a harmonic oscillator [$U(x) = \frac{1}{2} m (0.01x)^2$]. 
        The solid blue, dash-dotted red, and dotted green curves correspond to the incident electrons' energy fluctuations $C p_0 = 0.05$, $0.1$ and $0.2$ (a.u.) respectively.
        In each case $p_0=10^{-4}$ a.u.
        Results for a free (i.e., $C=0$, $U{=}0$) wavepacket (dashed black) are shown for comparison.
        \label{fig:trapping}}
\end{figure}

The same strategy can also be used to trap an atom; 
by setting $U(x){=}{-}U_{\text{eff}}(x)$ in Eq.~\eqref{eqn:calB} 
the environment will mimic the potential $U_{\text{eff}}(x)$. 
%The physics underlying the environments for enhancing tunneling and trapping are the same.
Figure.~\ref{fig:trapping} depicts simulation results for a hydrogen atom immersed in 
trapping environments with different standard deviations of the environmental force $Cp_0$.
As in the tunneling case, larger values for $Cp_0$ for a given $p_0$ cause additional heating. 
This deteriorates the trapping via purity losses and wavepacket spreading.
Nevertheless, one can see that for each of the
cases depicted the environmentally trapped wavepacket remains more spatially localized than the free wavepacket.
These results suggest that the optimal strategy for trapping a particle is to use jets with the smallest $Cp_0$ for which~\eqref{eqn:pgg} is satisfied. 
%Footnote? This condition is necessary to ensure the validity of the microscopic description.

It was shown in Ref.~\cite{lemeshko2013dissipative} that non-conservative forces between atoms can lead to binding,  
even when the potential interaction is repulsive.

%%%%%%%%%%%%%%%%%%%%%%%%%%%%%%%%%%%%%%%%%%%%%%%%%%%%%%%%%%%%%%%%%%%%%%%%%%

\section{Exotic applications}
We have demonstrated that non-conservative forces can effectively mimic desired conservative interactions, however, the utility of such forces is much wider. 
Non-conservative forces can also be used to obtain modifications $G(p)$ to the dispersion relationship~(\ref{eqn:Ehrenfest1}) --
note that such modifications cannot be implemented via conservative forces. 
We consider two applications for such modifications: 
tuning the effective mass of a quantum particle and emulating relativistic effects. 

\begin{figure}
		\includegraphics[width=1\hsize]{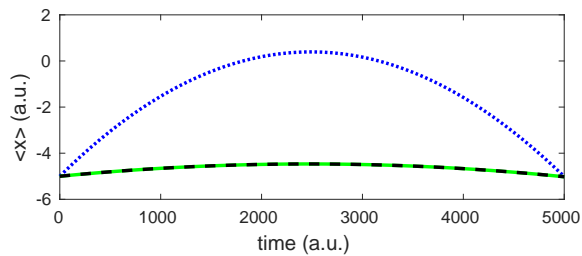}
        \caption{
        The expectation value of the position as a function of time for a hydrogen atom in a ramp potential $U(x) = 3.2 \times 10^{-3} x$ a.u.
        The dotted blue curve depicts the environment-free case. 
        The solid green curve depicts the atom in an environment \eqref{eqn:env3} [$M=10m$, $C=0.1$] engineered to give an effective mass ten times the proton mass $m$.
        The dynamics of the hydrogen atom with environmentally induced effective mass $M$ coincide with those of an environment-free particle of mass $M$ (dashed black).
        }
		\label{fig:ex3}
\end{figure}

Consider a quantum particle of mass $m$ in a potential $U(x)$. The particle will exhibit an effective mass $M$ when immersed in an environment 
%(obtained from the relations~\eqref{eqn:sol}) 
described by the dissipator 
$\mathcal{D}_{B}$ 
[as in Eq.~(\ref{eqn:dW})]
with
\begin{equation}  \label{eqn:env3}
B(p) = C \exp \left[ - \frac{i (m - M) p^2}{ 2mMC^2} \right].
\end{equation}
That is, the system dynamics will satisfy the constraints
\begin{equation}
 \frac{\dif \;}{\dif t} \langle \hat{x} \rangle 
    = \frac{1}{M} \langle \hat{p} \rangle, \quad 
\frac{\dif}{\dif t} \langle \hat{p} \rangle 
    = 
    - \left\langle \frac{\dif U(\hat{x})} {\dif \hat{x}} \right\rangle
    .   \label{eqn:Ehrenfest3}
\end{equation}
Figure~\ref{fig:ex3} depicts simulation results:  
the particle of mass $m$ in the environment \eqref{eqn:env3} evolves in excellent agreement with an environment-free particle of mass $M$. 

The effective mass approximation is ubiquitously used to describe the motion of a quantum particle in the periodic field of a solid. 
Recently, a negative effective mass was experimentally achieved \cite{khamehchi2017negative}. 
An atom interacting with the standing wave of a single photon in the cavity also acquires an effective mass \cite{larson2005effective}. 
We conjecture that environmentally induced mass can emerge for an atom elastically scattering off incoherent light seeded into a cavity.

%%%%%%%%%%%%%%%%%%%%%%%%%%%%%%%%%%%%%%%%%%%%%%%%%%%%%%%%%%%%%%%%%%%%%%%%%%

We now turn our attention to environmentally induced quasi-relativistic behaviour. 
Once again, consider a quantum particle of mass $m$ in a potential $U(x)$. 
Suppose we wish the system dynamics to satisfy the constraints
\begin{equation}
 \frac{\dif \;}{\dif t} \langle \hat{x} \rangle 
    = \left\langle \frac{\hat{p}}{\sqrt{m^2 + \hat{p}^2 / c^2}} \right\rangle, \quad 
\frac{\dif}{\dif t} \langle \hat{p} \rangle 
    = - \left\langle \frac{\dif U(\hat{x})} {\dif \hat{x}} \right\rangle. 
    \label{eqn:Ehrenfest4}  
\end{equation}
This can be achieved with an environment described by the dissipator 
$\mathcal{D}_{B}$ 
%[as in Eq.~(\ref{eqn:dW})]
with
\begin{equation}  \label{eqn:env4}
B(p) = C \exp \left[ \frac{i}{C^2} \left( \frac{p^2}{2m} - c \sqrt{m^2c^2 + p^2} \right)   \right].
\end{equation}

\begin{figure}
		\includegraphics[width=1\hsize]{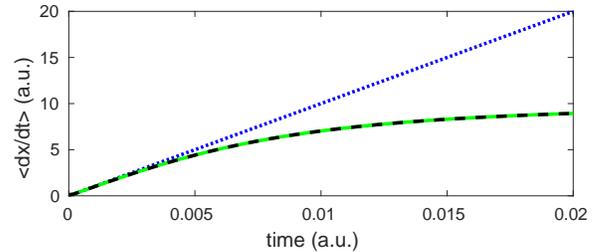}
        \caption{The expectation value of the velocity as a function of time for an electron in a ramp potential $U(x) = - 10^3 x$ a.u.
        The dotted blue curve depicts the environment-free case. 
        The solid green curve depicts the electron in an environment \eqref{eqn:env4} ($C=20$ a.u.) engineered to induce quasi-relativistic behaviour with the speed of light chosen to be 7\% of the speed of light in vacuum. 
        The environmentally induced quasi-relativistic behaviour coincides with that of an environment-free relativistic electron with Hamiltonian $\hat{H}_{\mathrm{rel}} = \sqrt{c^2\hat{p}^2 + c^4} + U(\hat{x})$; $c=10$ (dashed black).}
    	\label{fig:ex4}
\end{figure}

Figure~\ref{fig:ex4} depicts simulation results
confirming that the chosen environment induces quasi-relativistic behaviour for an arbitrarily small speed of light. In particular, the environment mimics the effect of time dilation 
as the particle velocity approaches the chosen speed of light.

The dispersion relation emerges as an effective description of the self-interaction of a bare quantum particle with a larger system with some characteristic symmetry. Generalizing the logic of Ref. \cite{larson2005effective}, we conjecture that tailoring the spectral transmission characteristics of a cavity 
and employing multi-color electromagnetic radiation with specific photon statistics 
should provide access to a large class of dispersion relations.

%%%%%%%%%%%%%%%%%%%%%%%%%%%%%%%%%%%%%%%%%%%%%%%%%%%%%%%%%%%%%%%%%%%%%%%%%%

\textit{Outlook.} 
Physicists, chemists, and engineers are increasingly looking for new ways to 
manipulate quantum systems -- non-conservative environments provide one such resource. 
We give a systematic approach 
for designing such environments to steer wavepackets along desired trajectories.
The method is demonstrated via several examples: 
enhancing quantum tunneling, 
trapping particles,
inducing effective mass, and 
emulating relativistic effects.
The proposed dissipators not only enhance desired quantum properties, 
they can be engineered to do so while preserving the purity of the underlying system.
A distinct feature of our method is that for a given $F$ and $G$, 
the resulting dynamics always satisfy 
Eqs.~\eqref{__problem_gen}, 
irrespective of the initial state.

Finally, note that $F(x)$ in Eq.~(\ref{eqn:Ehrenfest2}) is the sum of the potential force $-\dif U(x) /\dif x$ 
and the environmentally induced forces $f_k(x)$ [Eq.~\eqref{eqn:F_k}]. 
Despite being of different physical origins [Eqs.~(\ref{eqn:H}) and (\ref{eqn:A}) respectively], 
these forces contribute to $F(x)$ on an equal footing.
This observation may help shed light on the discussion regarding the entropic interpretation of the gravitational force \cite{verlinde2011origin}. 
In this regard, it would be beneficial to find a dynamical signature that could efficiently discriminate between potential and statistical interactions.

\textit{Acknowledgments.}
S.L.V. was supported by the Australian Research Council
(DP130104510). D.I.B., R.C. respectively acknowledge financial support 
from NSF CHE 1464569 and DOE DE-FG-02-ER-15344.  
T. S.  thanks the National Science Foundation (Award No. CHE-1465201) for support. 
D. I. B. is also supported by Humboldt Research Fellowship for Experienced Researchers and AFOSR Young Investigator Research Program (No. FA9550-16-1-0254).

\bibliography{qre}

\begin{thebibliography}{47}%
\makeatletter
\providecommand \@ifxundefined [1]{%
 \@ifx{#1\undefined}
}%
\providecommand \@ifnum [1]{%
 \ifnum #1\expandafter \@firstoftwo
 \else \expandafter \@secondoftwo
 \fi
}%
\providecommand \@ifx [1]{%
 \ifx #1\expandafter \@firstoftwo
 \else \expandafter \@secondoftwo
 \fi
}%
\providecommand \natexlab [1]{#1}%
\providecommand \enquote  [1]{``#1''}%
\providecommand \bibnamefont  [1]{#1}%
\providecommand \bibfnamefont [1]{#1}%
\providecommand \citenamefont [1]{#1}%
\providecommand \href@noop [0]{\@secondoftwo}%
\providecommand \href [0]{\begingroup \@sanitize@url \@href}%
\providecommand \@href[1]{\@@startlink{#1}\@@href}%
\providecommand \@@href[1]{\endgroup#1\@@endlink}%
\providecommand \@sanitize@url [0]{\catcode `\\12\catcode `\$12\catcode
  `\&12\catcode `\#12\catcode `\^12\catcode `\_12\catcode `\%12\relax}%
\providecommand \@@startlink[1]{}%
\providecommand \@@endlink[0]{}%
\providecommand \url  [0]{\begingroup\@sanitize@url \@url }%
\providecommand \@url [1]{\endgroup\@href {#1}{\urlprefix }}%
\providecommand \urlprefix  [0]{URL }%
\providecommand \Eprint [0]{\href }%
\providecommand \doibase [0]{http://dx.doi.org/}%
\providecommand \selectlanguage [0]{\@gobble}%
\providecommand \bibinfo  [0]{\@secondoftwo}%
\providecommand \bibfield  [0]{\@secondoftwo}%
\providecommand \translation [1]{[#1]}%
\providecommand \BibitemOpen [0]{}%
\providecommand \bibitemStop [0]{}%
\providecommand \bibitemNoStop [0]{.\EOS\space}%
\providecommand \EOS [0]{\spacefactor3000\relax}%
\providecommand \BibitemShut  [1]{\csname bibitem#1\endcsname}%
\let\auto@bib@innerbib\@empty
%</preamble>
\bibitem [{\citenamefont {Poyatos}\ \emph {et~al.}(1996)\citenamefont
  {Poyatos}, \citenamefont {Cirac},\ and\ \citenamefont
  {Zoller}}]{poyatos1996quantum}%
  \BibitemOpen
  \bibfield  {author} {\bibinfo {author} {\bibfnamefont {J.}~\bibnamefont
  {Poyatos}}, \bibinfo {author} {\bibfnamefont {J.}~\bibnamefont {Cirac}}, \
  and\ \bibinfo {author} {\bibfnamefont {P.}~\bibnamefont {Zoller}},\
  }\href@noop {} {\bibfield  {journal} {\bibinfo  {journal} {Phys. Rev. Lett.}\
  }\textbf {\bibinfo {volume} {77}},\ \bibinfo {pages} {4728} (\bibinfo {year}
  {1996})}\BibitemShut {NoStop}%
\bibitem [{\citenamefont {Verstraete}\ \emph {et~al.}(2009)\citenamefont
  {Verstraete}, \citenamefont {Wolf},\ and\ \citenamefont
  {Cirac}}]{verstraete2009quantum}%
  \BibitemOpen
  \bibfield  {author} {\bibinfo {author} {\bibfnamefont {F.}~\bibnamefont
  {Verstraete}}, \bibinfo {author} {\bibfnamefont {M.~M.}\ \bibnamefont
  {Wolf}}, \ and\ \bibinfo {author} {\bibfnamefont {J.~I.}\ \bibnamefont
  {Cirac}},\ }\href@noop {} {\bibfield  {journal} {\bibinfo  {journal} {Nature
  Physics}\ }\textbf {\bibinfo {volume} {5}},\ \bibinfo {pages} {633} (\bibinfo
  {year} {2009})}\BibitemShut {NoStop}%
\bibitem [{\citenamefont {Fedortchenko}\ \emph {et~al.}(2014)\citenamefont
  {Fedortchenko}, \citenamefont {Keller}, \citenamefont {Coudreau},\ and\
  \citenamefont {Milman}}]{fedortchenko2014finite}%
  \BibitemOpen
  \bibfield  {author} {\bibinfo {author} {\bibfnamefont {S.}~\bibnamefont
  {Fedortchenko}}, \bibinfo {author} {\bibfnamefont {A.}~\bibnamefont
  {Keller}}, \bibinfo {author} {\bibfnamefont {T.}~\bibnamefont {Coudreau}}, \
  and\ \bibinfo {author} {\bibfnamefont {P.}~\bibnamefont {Milman}},\
  }\href@noop {} {\bibfield  {journal} {\bibinfo  {journal} {Phy. Rev. A}\
  }\textbf {\bibinfo {volume} {90}},\ \bibinfo {pages} {042103} (\bibinfo
  {year} {2014})}\BibitemShut {NoStop}%
\bibitem [{\citenamefont {Kurizki}\ \emph {et~al.}(2015)\citenamefont
  {Kurizki}, \citenamefont {Shahmoon},\ and\ \citenamefont
  {Zwick}}]{kurizki2015thermal}%
  \BibitemOpen
  \bibfield  {author} {\bibinfo {author} {\bibfnamefont {G.}~\bibnamefont
  {Kurizki}}, \bibinfo {author} {\bibfnamefont {E.}~\bibnamefont {Shahmoon}}, \
  and\ \bibinfo {author} {\bibfnamefont {A.}~\bibnamefont {Zwick}},\ }\href
  {http://stacks.iop.org/1402-4896/90/i=12/a=128002} {\bibfield  {journal}
  {\bibinfo  {journal} {Physica Scripta}\ }\textbf {\bibinfo {volume} {90}},\
  \bibinfo {pages} {128002} (\bibinfo {year} {2015})}\BibitemShut {NoStop}%
\bibitem [{\citenamefont {Kienzler}\ \emph {et~al.}(2015)\citenamefont
  {Kienzler}, \citenamefont {Lo}, \citenamefont {Keitch}, \citenamefont
  {de~Clercq}, \citenamefont {Leupold}, \citenamefont {Lindenfelser},
  \citenamefont {Marinelli}, \citenamefont {Negnevitsky},\ and\ \citenamefont
  {Home}}]{kienzler2015quantum}%
  \BibitemOpen
  \bibfield  {author} {\bibinfo {author} {\bibfnamefont {D.}~\bibnamefont
  {Kienzler}}, \bibinfo {author} {\bibfnamefont {H.-Y.}\ \bibnamefont {Lo}},
  \bibinfo {author} {\bibfnamefont {B.}~\bibnamefont {Keitch}}, \bibinfo
  {author} {\bibfnamefont {L.}~\bibnamefont {de~Clercq}}, \bibinfo {author}
  {\bibfnamefont {F.}~\bibnamefont {Leupold}}, \bibinfo {author} {\bibfnamefont
  {F.}~\bibnamefont {Lindenfelser}}, \bibinfo {author} {\bibfnamefont
  {M.}~\bibnamefont {Marinelli}}, \bibinfo {author} {\bibfnamefont
  {V.}~\bibnamefont {Negnevitsky}}, \ and\ \bibinfo {author} {\bibfnamefont
  {J.}~\bibnamefont {Home}},\ }\href@noop {} {\bibfield  {journal} {\bibinfo
  {journal} {Science}\ }\textbf {\bibinfo {volume} {347}},\ \bibinfo {pages}
  {53} (\bibinfo {year} {2015})}\BibitemShut {NoStop}%
\bibitem [{\citenamefont {Pan}\ \emph {et~al.}(2016)\citenamefont {Pan},
  \citenamefont {Ugrinovskii},\ and\ \citenamefont {James}}]{pan2016ground}%
  \BibitemOpen
  \bibfield  {author} {\bibinfo {author} {\bibfnamefont {Y.}~\bibnamefont
  {Pan}}, \bibinfo {author} {\bibfnamefont {V.}~\bibnamefont {Ugrinovskii}}, \
  and\ \bibinfo {author} {\bibfnamefont {M.~R.}\ \bibnamefont {James}},\ }\href
  {\doibase http://dx.doi.org/10.1016/j.automatica.2015.11.041} {\bibfield
  {journal} {\bibinfo  {journal} {Automatica}\ }\textbf {\bibinfo {volume}
  {65}},\ \bibinfo {pages} {147 } (\bibinfo {year} {2016})}\BibitemShut
  {NoStop}%
\bibitem [{\citenamefont {Rouchon}(2014)}]{rouchon2014models}%
  \BibitemOpen
  \bibfield  {author} {\bibinfo {author} {\bibfnamefont {P.}~\bibnamefont
  {Rouchon}},\ }\href@noop {} {\bibfield  {journal} {\bibinfo  {journal}
  {arXiv:1407.7810}\ } (\bibinfo {year} {2014})}\BibitemShut {NoStop}%
\bibitem [{\citenamefont {Metelmann}\ and\ \citenamefont
  {Clerk}(2014)}]{metelmann2014quantum}%
  \BibitemOpen
  \bibfield  {author} {\bibinfo {author} {\bibfnamefont {A.}~\bibnamefont
  {Metelmann}}\ and\ \bibinfo {author} {\bibfnamefont {A.}~\bibnamefont
  {Clerk}},\ }\href@noop {} {\bibfield  {journal} {\bibinfo  {journal} {Phys.
  Rev. Lett.}\ }\textbf {\bibinfo {volume} {112}},\ \bibinfo {pages} {133904}
  (\bibinfo {year} {2014})}\BibitemShut {NoStop}%
\bibitem [{\citenamefont {Metelmann}\ and\ \citenamefont
  {Clerk}(2015)}]{metelmann2015nonreciprocal}%
  \BibitemOpen
  \bibfield  {author} {\bibinfo {author} {\bibfnamefont {A.}~\bibnamefont
  {Metelmann}}\ and\ \bibinfo {author} {\bibfnamefont {A.~A.}\ \bibnamefont
  {Clerk}},\ }\href {\doibase 10.1103/PhysRevX.5.021025} {\bibfield  {journal}
  {\bibinfo  {journal} {Phys. Rev. X}\ }\textbf {\bibinfo {volume} {5}},\
  \bibinfo {pages} {021025} (\bibinfo {year} {2015})}\BibitemShut {NoStop}%
\bibitem [{\citenamefont {Metelmann}\ and\ \citenamefont
  {Clerk}(2017)}]{metelmann2017nonreciprocal}%
  \BibitemOpen
  \bibfield  {author} {\bibinfo {author} {\bibfnamefont {A.}~\bibnamefont
  {Metelmann}}\ and\ \bibinfo {author} {\bibfnamefont {A.}~\bibnamefont
  {Clerk}},\ }\href@noop {} {\bibfield  {journal} {\bibinfo  {journal} {Phys.
  Rev. A}\ }\textbf {\bibinfo {volume} {95}},\ \bibinfo {pages} {013837}
  (\bibinfo {year} {2017})}\BibitemShut {NoStop}%
\bibitem [{\citenamefont {Miranowicz}\ \emph {et~al.}(2014)\citenamefont
  {Miranowicz}, \citenamefont {Bajer}, \citenamefont {Paprzycka}, \citenamefont
  {Liu}, \citenamefont {Zagoskin},\ and\ \citenamefont
  {Nori}}]{miranowicz2014state}%
  \BibitemOpen
  \bibfield  {author} {\bibinfo {author} {\bibfnamefont {A.}~\bibnamefont
  {Miranowicz}}, \bibinfo {author} {\bibfnamefont {J.}~\bibnamefont {Bajer}},
  \bibinfo {author} {\bibfnamefont {M.}~\bibnamefont {Paprzycka}}, \bibinfo
  {author} {\bibfnamefont {Y.-x.}\ \bibnamefont {Liu}}, \bibinfo {author}
  {\bibfnamefont {A.~M.}\ \bibnamefont {Zagoskin}}, \ and\ \bibinfo {author}
  {\bibfnamefont {F.}~\bibnamefont {Nori}},\ }\href {\doibase
  10.1103/PhysRevA.90.033831} {\bibfield  {journal} {\bibinfo  {journal} {Phys.
  Rev. A}\ }\textbf {\bibinfo {volume} {90}},\ \bibinfo {pages} {033831}
  (\bibinfo {year} {2014})}\BibitemShut {NoStop}%
\bibitem [{\citenamefont {Zhdanov}\ and\ \citenamefont
  {Seideman}(2015)}]{Zhdanov2015quantum}%
  \BibitemOpen
  \bibfield  {author} {\bibinfo {author} {\bibfnamefont {D.~V.}\ \bibnamefont
  {Zhdanov}}\ and\ \bibinfo {author} {\bibfnamefont {T.}~\bibnamefont
  {Seideman}},\ }\href@noop {} {\bibfield  {journal} {\bibinfo  {journal}
  {arXiv:1508.04481}\ } (\bibinfo {year} {2015})}\BibitemShut {NoStop}%
\bibitem [{\citenamefont {Lemeshko}\ and\ \citenamefont
  {Weimer}(2013)}]{lemeshko2013dissipative}%
  \BibitemOpen
  \bibfield  {author} {\bibinfo {author} {\bibfnamefont {M.}~\bibnamefont
  {Lemeshko}}\ and\ \bibinfo {author} {\bibfnamefont {H.}~\bibnamefont
  {Weimer}},\ }\href@noop {} {\bibfield  {journal} {\bibinfo  {journal} {Nature
  Communications}\ }\textbf {\bibinfo {volume} {4}} (\bibinfo {year}
  {2013})}\BibitemShut {NoStop}%
\bibitem [{\citenamefont {W{\"u}ster}(2017)}]{wuster2017quantum}%
  \BibitemOpen
  \bibfield  {author} {\bibinfo {author} {\bibfnamefont {S.}~\bibnamefont
  {W{\"u}ster}},\ }\href@noop {} {\bibfield  {journal} {\bibinfo  {journal}
  {Phys. Rev. Lett.}\ }\textbf {\bibinfo {volume} {119}},\ \bibinfo {pages}
  {013001} (\bibinfo {year} {2017})}\BibitemShut {NoStop}%
\bibitem [{\citenamefont {Kaczmarczyk}\ \emph {et~al.}(2016)\citenamefont
  {Kaczmarczyk}, \citenamefont {Weimer},\ and\ \citenamefont
  {Lemeshko}}]{kaczmarczyk2016dissipative}%
  \BibitemOpen
  \bibfield  {author} {\bibinfo {author} {\bibfnamefont {J.}~\bibnamefont
  {Kaczmarczyk}}, \bibinfo {author} {\bibfnamefont {H.}~\bibnamefont {Weimer}},
  \ and\ \bibinfo {author} {\bibfnamefont {M.}~\bibnamefont {Lemeshko}},\
  }\href@noop {} {\bibfield  {journal} {\bibinfo  {journal} {New J. Phys.}\
  }\textbf {\bibinfo {volume} {18}},\ \bibinfo {pages} {093042} (\bibinfo
  {year} {2016})}\BibitemShut {NoStop}%
\bibitem [{\citenamefont {Weimer}(2016)}]{weimer2016tailored}%
  \BibitemOpen
  \bibfield  {author} {\bibinfo {author} {\bibfnamefont {H.}~\bibnamefont
  {Weimer}},\ }\href@noop {} {\bibfield  {journal} {\bibinfo  {journal} {J.
  Phys. B}\ }\textbf {\bibinfo {volume} {50}},\ \bibinfo {pages} {024001}
  (\bibinfo {year} {2016})}\BibitemShut {NoStop}%
\bibitem [{\citenamefont {Overbeck}\ \emph {et~al.}(2017)\citenamefont
  {Overbeck}, \citenamefont {Maghrebi}, \citenamefont {Gorshkov},\ and\
  \citenamefont {Weimer}}]{overbeck2017multicritical}%
  \BibitemOpen
  \bibfield  {author} {\bibinfo {author} {\bibfnamefont {V.~R.}\ \bibnamefont
  {Overbeck}}, \bibinfo {author} {\bibfnamefont {M.~F.}\ \bibnamefont
  {Maghrebi}}, \bibinfo {author} {\bibfnamefont {A.~V.}\ \bibnamefont
  {Gorshkov}}, \ and\ \bibinfo {author} {\bibfnamefont {H.}~\bibnamefont
  {Weimer}},\ }\href@noop {} {\bibfield  {journal} {\bibinfo  {journal} {Phys.
  Rev. A}\ }\textbf {\bibinfo {volume} {95}},\ \bibinfo {pages} {042133}
  (\bibinfo {year} {2017})}\BibitemShut {NoStop}%
\bibitem [{\citenamefont {Albert}\ \emph {et~al.}(2016)\citenamefont {Albert},
  \citenamefont {Shu}, \citenamefont {Krastanov}, \citenamefont {Shen},
  \citenamefont {Liu}, \citenamefont {Yang}, \citenamefont {Schoelkopf},
  \citenamefont {Mirrahimi}, \citenamefont {Devoret},\ and\ \citenamefont
  {Jiang}}]{albert2016holonomic}%
  \BibitemOpen
  \bibfield  {author} {\bibinfo {author} {\bibfnamefont {V.~V.}\ \bibnamefont
  {Albert}}, \bibinfo {author} {\bibfnamefont {C.}~\bibnamefont {Shu}},
  \bibinfo {author} {\bibfnamefont {S.}~\bibnamefont {Krastanov}}, \bibinfo
  {author} {\bibfnamefont {C.}~\bibnamefont {Shen}}, \bibinfo {author}
  {\bibfnamefont {R.-B.}\ \bibnamefont {Liu}}, \bibinfo {author} {\bibfnamefont
  {Z.-B.}\ \bibnamefont {Yang}}, \bibinfo {author} {\bibfnamefont {R.~J.}\
  \bibnamefont {Schoelkopf}}, \bibinfo {author} {\bibfnamefont
  {M.}~\bibnamefont {Mirrahimi}}, \bibinfo {author} {\bibfnamefont {M.~H.}\
  \bibnamefont {Devoret}}, \ and\ \bibinfo {author} {\bibfnamefont
  {L.}~\bibnamefont {Jiang}},\ }\href@noop {} {\bibfield  {journal} {\bibinfo
  {journal} {Phys. Rev. Lett.}\ }\textbf {\bibinfo {volume} {116}},\ \bibinfo
  {pages} {140502} (\bibinfo {year} {2016})}\BibitemShut {NoStop}%
\bibitem [{\citenamefont {Ticozzi}\ and\ \citenamefont
  {Viola}(2017)}]{ticozzi2017quantum}%
  \BibitemOpen
  \bibfield  {author} {\bibinfo {author} {\bibfnamefont {F.}~\bibnamefont
  {Ticozzi}}\ and\ \bibinfo {author} {\bibfnamefont {L.}~\bibnamefont
  {Viola}},\ }\href {http://stacks.iop.org/2058-9565/2/i=3/a=034001} {\bibfield
   {journal} {\bibinfo  {journal} {Quantum Science and Technology}\ }\textbf
  {\bibinfo {volume} {2}},\ \bibinfo {pages} {034001} (\bibinfo {year}
  {2017})}\BibitemShut {NoStop}%
\bibitem [{\citenamefont {Arenz}\ \emph {et~al.}(2016)\citenamefont {Arenz},
  \citenamefont {Burgarth}, \citenamefont {Facchi}, \citenamefont
  {Giovannetti}, \citenamefont {Nakazato}, \citenamefont {Pascazio},\ and\
  \citenamefont {Yuasa}}]{arenz2016universal}%
  \BibitemOpen
  \bibfield  {author} {\bibinfo {author} {\bibfnamefont {C.}~\bibnamefont
  {Arenz}}, \bibinfo {author} {\bibfnamefont {D.}~\bibnamefont {Burgarth}},
  \bibinfo {author} {\bibfnamefont {P.}~\bibnamefont {Facchi}}, \bibinfo
  {author} {\bibfnamefont {V.}~\bibnamefont {Giovannetti}}, \bibinfo {author}
  {\bibfnamefont {H.}~\bibnamefont {Nakazato}}, \bibinfo {author}
  {\bibfnamefont {S.}~\bibnamefont {Pascazio}}, \ and\ \bibinfo {author}
  {\bibfnamefont {K.}~\bibnamefont {Yuasa}},\ }\href@noop {} {\bibfield
  {journal} {\bibinfo  {journal} {Phys. Rev. A}\ }\textbf {\bibinfo {volume}
  {93}},\ \bibinfo {pages} {062308} (\bibinfo {year} {2016})}\BibitemShut
  {NoStop}%
\bibitem [{\citenamefont {Arenz}\ \emph {et~al.}(2017)\citenamefont {Arenz},
  \citenamefont {Burgarth}, \citenamefont {Giovannetti}, \citenamefont
  {Nakazato},\ and\ \citenamefont {Yuasa}}]{arenz2017lindbladian}%
  \BibitemOpen
  \bibfield  {author} {\bibinfo {author} {\bibfnamefont {C.}~\bibnamefont
  {Arenz}}, \bibinfo {author} {\bibfnamefont {D.}~\bibnamefont {Burgarth}},
  \bibinfo {author} {\bibfnamefont {V.}~\bibnamefont {Giovannetti}}, \bibinfo
  {author} {\bibfnamefont {H.}~\bibnamefont {Nakazato}}, \ and\ \bibinfo
  {author} {\bibfnamefont {K.}~\bibnamefont {Yuasa}},\ }\href@noop {}
  {\bibfield  {journal} {\bibinfo  {journal} {Quantum Science and Technology}\
  }\textbf {\bibinfo {volume} {2}},\ \bibinfo {pages} {024001} (\bibinfo {year}
  {2017})}\BibitemShut {NoStop}%
\bibitem [{\citenamefont {Cheng}\ \emph {et~al.}(2016)\citenamefont {Cheng},
  \citenamefont {Zhang}, \citenamefont {Zhou},\ and\ \citenamefont
  {Zhang}}]{cheng2016preservation}%
  \BibitemOpen
  \bibfield  {author} {\bibinfo {author} {\bibfnamefont {J.}~\bibnamefont
  {Cheng}}, \bibinfo {author} {\bibfnamefont {W.-Z.}\ \bibnamefont {Zhang}},
  \bibinfo {author} {\bibfnamefont {L.}~\bibnamefont {Zhou}}, \ and\ \bibinfo
  {author} {\bibfnamefont {W.}~\bibnamefont {Zhang}},\ }\href@noop {}
  {\bibfield  {journal} {\bibinfo  {journal} {Scientific reports}\ }\textbf
  {\bibinfo {volume} {6}},\ \bibinfo {pages} {23678} (\bibinfo {year}
  {2016})}\BibitemShut {NoStop}%
\bibitem [{\citenamefont {Liu}\ \emph {et~al.}(2016)\citenamefont {Liu},
  \citenamefont {Shankar}, \citenamefont {Ofek}, \citenamefont {Hatridge},
  \citenamefont {Narla}, \citenamefont {Sliwa}, \citenamefont {Frunzio},
  \citenamefont {Schoelkopf},\ and\ \citenamefont
  {Devoret}}]{liu2016comparing}%
  \BibitemOpen
  \bibfield  {author} {\bibinfo {author} {\bibfnamefont {Y.}~\bibnamefont
  {Liu}}, \bibinfo {author} {\bibfnamefont {S.}~\bibnamefont {Shankar}},
  \bibinfo {author} {\bibfnamefont {N.}~\bibnamefont {Ofek}}, \bibinfo {author}
  {\bibfnamefont {M.}~\bibnamefont {Hatridge}}, \bibinfo {author}
  {\bibfnamefont {A.}~\bibnamefont {Narla}}, \bibinfo {author} {\bibfnamefont
  {K.}~\bibnamefont {Sliwa}}, \bibinfo {author} {\bibfnamefont
  {L.}~\bibnamefont {Frunzio}}, \bibinfo {author} {\bibfnamefont {R.~J.}\
  \bibnamefont {Schoelkopf}}, \ and\ \bibinfo {author} {\bibfnamefont {M.~H.}\
  \bibnamefont {Devoret}},\ }\href@noop {} {\bibfield  {journal} {\bibinfo
  {journal} {Physical Review X}\ }\textbf {\bibinfo {volume} {6}},\ \bibinfo
  {pages} {011022} (\bibinfo {year} {2016})}\BibitemShut {NoStop}%
\bibitem [{\citenamefont {Zippilli}\ \emph {et~al.}(2015)\citenamefont
  {Zippilli}, \citenamefont {Li},\ and\ \citenamefont
  {Vitali}}]{zippilli2015steady}%
  \BibitemOpen
  \bibfield  {author} {\bibinfo {author} {\bibfnamefont {S.}~\bibnamefont
  {Zippilli}}, \bibinfo {author} {\bibfnamefont {J.}~\bibnamefont {Li}}, \ and\
  \bibinfo {author} {\bibfnamefont {D.}~\bibnamefont {Vitali}},\ }\href
  {\doibase 10.1103/PhysRevA.92.032319} {\bibfield  {journal} {\bibinfo
  {journal} {Phys. Rev. A}\ }\textbf {\bibinfo {volume} {92}},\ \bibinfo
  {pages} {032319} (\bibinfo {year} {2015})}\BibitemShut {NoStop}%
\bibitem [{\citenamefont {Yang}\ \emph {et~al.}(2015)\citenamefont {Yang},
  \citenamefont {An}, \citenamefont {Yang},\ and\ \citenamefont
  {Li}}]{yang2015generation}%
  \BibitemOpen
  \bibfield  {author} {\bibinfo {author} {\bibfnamefont {C.-J.}\ \bibnamefont
  {Yang}}, \bibinfo {author} {\bibfnamefont {J.-H.}\ \bibnamefont {An}},
  \bibinfo {author} {\bibfnamefont {W.}~\bibnamefont {Yang}}, \ and\ \bibinfo
  {author} {\bibfnamefont {Y.}~\bibnamefont {Li}},\ }\href {\doibase
  10.1103/PhysRevA.92.062311} {\bibfield  {journal} {\bibinfo  {journal} {Phys.
  Rev. A}\ }\textbf {\bibinfo {volume} {92}},\ \bibinfo {pages} {062311}
  (\bibinfo {year} {2015})}\BibitemShut {NoStop}%
\bibitem [{\citenamefont {Mirza}(2015)}]{mirza2015controlling}%
  \BibitemOpen
  \bibfield  {author} {\bibinfo {author} {\bibfnamefont {I.~M.}\ \bibnamefont
  {Mirza}},\ }\href@noop {} {\bibfield  {journal} {\bibinfo  {journal} {Journal
  of Modern Optics}\ }\textbf {\bibinfo {volume} {62}},\ \bibinfo {pages}
  {1048} (\bibinfo {year} {2015})}\BibitemShut {NoStop}%
\bibitem [{\citenamefont {Arenz}\ \emph {et~al.}(2013)\citenamefont {Arenz},
  \citenamefont {Cormick}, \citenamefont {Vitali},\ and\ \citenamefont
  {Morigi}}]{arenz2013generation}%
  \BibitemOpen
  \bibfield  {author} {\bibinfo {author} {\bibfnamefont {C.}~\bibnamefont
  {Arenz}}, \bibinfo {author} {\bibfnamefont {C.}~\bibnamefont {Cormick}},
  \bibinfo {author} {\bibfnamefont {D.}~\bibnamefont {Vitali}}, \ and\ \bibinfo
  {author} {\bibfnamefont {G.}~\bibnamefont {Morigi}},\ }\href@noop {}
  {\bibfield  {journal} {\bibinfo  {journal} {J. Phys. B}\ }\textbf {\bibinfo
  {volume} {46}},\ \bibinfo {pages} {224001} (\bibinfo {year}
  {2013})}\BibitemShut {NoStop}%
\bibitem [{\citenamefont {Kronwald}\ \emph {et~al.}(2013)\citenamefont
  {Kronwald}, \citenamefont {Marquardt},\ and\ \citenamefont
  {Clerk}}]{kronwald2013arbitrarily}%
  \BibitemOpen
  \bibfield  {author} {\bibinfo {author} {\bibfnamefont {A.}~\bibnamefont
  {Kronwald}}, \bibinfo {author} {\bibfnamefont {F.}~\bibnamefont {Marquardt}},
  \ and\ \bibinfo {author} {\bibfnamefont {A.~A.}\ \bibnamefont {Clerk}},\
  }\href {\doibase 10.1103/PhysRevA.88.063833} {\bibfield  {journal} {\bibinfo
  {journal} {Phys. Rev. A}\ }\textbf {\bibinfo {volume} {88}},\ \bibinfo
  {pages} {063833} (\bibinfo {year} {2013})}\BibitemShut {NoStop}%
\bibitem [{\citenamefont {Woolley}\ and\ \citenamefont
  {Clerk}(2014)}]{woolley2014two}%
  \BibitemOpen
  \bibfield  {author} {\bibinfo {author} {\bibfnamefont {M.~J.}\ \bibnamefont
  {Woolley}}\ and\ \bibinfo {author} {\bibfnamefont {A.~A.}\ \bibnamefont
  {Clerk}},\ }\href {\doibase 10.1103/PhysRevA.89.063805} {\bibfield  {journal}
  {\bibinfo  {journal} {Phys. Rev. A}\ }\textbf {\bibinfo {volume} {89}},\
  \bibinfo {pages} {063805} (\bibinfo {year} {2014})}\BibitemShut {NoStop}%
\bibitem [{\citenamefont {Grimsmo}\ \emph {et~al.}(2016)\citenamefont
  {Grimsmo}, \citenamefont {Qassemi}, \citenamefont {Reulet},\ and\
  \citenamefont {Blais}}]{grimsmo2016quantum}%
  \BibitemOpen
  \bibfield  {author} {\bibinfo {author} {\bibfnamefont {A.~L.}\ \bibnamefont
  {Grimsmo}}, \bibinfo {author} {\bibfnamefont {F.}~\bibnamefont {Qassemi}},
  \bibinfo {author} {\bibfnamefont {B.}~\bibnamefont {Reulet}}, \ and\ \bibinfo
  {author} {\bibfnamefont {A.}~\bibnamefont {Blais}},\ }\href@noop {}
  {\bibfield  {journal} {\bibinfo  {journal} {Phys. Rev. Lett.}\ }\textbf
  {\bibinfo {volume} {116}},\ \bibinfo {pages} {043602} (\bibinfo {year}
  {2016})}\BibitemShut {NoStop}%
\bibitem [{\citenamefont {Koga}\ and\ \citenamefont
  {Yamamoto}(2012)}]{koga2012dissipation}%
  \BibitemOpen
  \bibfield  {author} {\bibinfo {author} {\bibfnamefont {K.}~\bibnamefont
  {Koga}}\ and\ \bibinfo {author} {\bibfnamefont {N.}~\bibnamefont
  {Yamamoto}},\ }\href {\doibase 10.1103/PhysRevA.85.022103} {\bibfield
  {journal} {\bibinfo  {journal} {Phys. Rev. A}\ }\textbf {\bibinfo {volume}
  {85}},\ \bibinfo {pages} {022103} (\bibinfo {year} {2012})}\BibitemShut
  {NoStop}%
\bibitem [{\citenamefont {Holland}\ \emph {et~al.}(2015)\citenamefont
  {Holland}, \citenamefont {Vlastakis}, \citenamefont {Heeres}, \citenamefont
  {Reagor}, \citenamefont {Vool}, \citenamefont {Leghtas}, \citenamefont
  {Frunzio}, \citenamefont {Kirchmair}, \citenamefont {Devoret}, \citenamefont
  {Mirrahimi},\ and\ \citenamefont {Schoelkopf}}]{holland2015single}%
  \BibitemOpen
  \bibfield  {author} {\bibinfo {author} {\bibfnamefont {E.~T.}\ \bibnamefont
  {Holland}}, \bibinfo {author} {\bibfnamefont {B.}~\bibnamefont {Vlastakis}},
  \bibinfo {author} {\bibfnamefont {R.~W.}\ \bibnamefont {Heeres}}, \bibinfo
  {author} {\bibfnamefont {M.~J.}\ \bibnamefont {Reagor}}, \bibinfo {author}
  {\bibfnamefont {U.}~\bibnamefont {Vool}}, \bibinfo {author} {\bibfnamefont
  {Z.}~\bibnamefont {Leghtas}}, \bibinfo {author} {\bibfnamefont
  {L.}~\bibnamefont {Frunzio}}, \bibinfo {author} {\bibfnamefont
  {G.}~\bibnamefont {Kirchmair}}, \bibinfo {author} {\bibfnamefont {M.~H.}\
  \bibnamefont {Devoret}}, \bibinfo {author} {\bibfnamefont {M.}~\bibnamefont
  {Mirrahimi}}, \ and\ \bibinfo {author} {\bibfnamefont {R.~J.}\ \bibnamefont
  {Schoelkopf}},\ }\href {\doibase 10.1103/PhysRevLett.115.180501} {\bibfield
  {journal} {\bibinfo  {journal} {Phys. Rev. Lett.}\ }\textbf {\bibinfo
  {volume} {115}},\ \bibinfo {pages} {180501} (\bibinfo {year}
  {2015})}\BibitemShut {NoStop}%
\bibitem [{\citenamefont {Asjad}\ and\ \citenamefont
  {Vitali}(2014)}]{asjad2014reservoir}%
  \BibitemOpen
  \bibfield  {author} {\bibinfo {author} {\bibfnamefont {M.}~\bibnamefont
  {Asjad}}\ and\ \bibinfo {author} {\bibfnamefont {D.}~\bibnamefont {Vitali}},\
  }\href {http://stacks.iop.org/0953-4075/47/i=4/a=045502} {\bibfield
  {journal} {\bibinfo  {journal} {J. Phys. B}\ }\textbf {\bibinfo {volume}
  {47}},\ \bibinfo {pages} {045502} (\bibinfo {year} {2014})}\BibitemShut
  {NoStop}%
\bibitem [{\citenamefont {Chestnov}\ \emph {et~al.}(2016)\citenamefont
  {Chestnov}, \citenamefont {Demirchyan}, \citenamefont {Alodjants},
  \citenamefont {Rubo},\ and\ \citenamefont {Kavokin}}]{chestnov2016permanent}%
  \BibitemOpen
  \bibfield  {author} {\bibinfo {author} {\bibfnamefont {I.~Y.}\ \bibnamefont
  {Chestnov}}, \bibinfo {author} {\bibfnamefont {S.~S.}\ \bibnamefont
  {Demirchyan}}, \bibinfo {author} {\bibfnamefont {A.~P.}\ \bibnamefont
  {Alodjants}}, \bibinfo {author} {\bibfnamefont {Y.~G.}\ \bibnamefont {Rubo}},
  \ and\ \bibinfo {author} {\bibfnamefont {A.~V.}\ \bibnamefont {Kavokin}},\
  }\href@noop {} {\bibfield  {journal} {\bibinfo  {journal} {Scientific
  reports}\ }\textbf {\bibinfo {volume} {6}},\ \bibinfo {pages} {19551}
  (\bibinfo {year} {2016})}\BibitemShut {NoStop}%
\bibitem [{\citenamefont {Bondar}\ \emph {et~al.}(2012)\citenamefont {Bondar},
  \citenamefont {Cabrera}, \citenamefont {Lompay}, \citenamefont {Ivanov},\
  and\ \citenamefont {Rabitz}}]{Bondar2011c}%
  \BibitemOpen
  \bibfield  {author} {\bibinfo {author} {\bibfnamefont {D.~I.}\ \bibnamefont
  {Bondar}}, \bibinfo {author} {\bibfnamefont {R.}~\bibnamefont {Cabrera}},
  \bibinfo {author} {\bibfnamefont {R.~R.}\ \bibnamefont {Lompay}}, \bibinfo
  {author} {\bibfnamefont {M.~Y.}\ \bibnamefont {Ivanov}}, \ and\ \bibinfo
  {author} {\bibfnamefont {H.~A.}\ \bibnamefont {Rabitz}},\ }\href {\doibase
  10.1103/PhysRevLett.109.190403} {\bibfield  {journal} {\bibinfo  {journal}
  {Phys. Rev. Lett.}\ }\textbf {\bibinfo {volume} {109}},\ \bibinfo {pages}
  {190403} (\bibinfo {year} {2012})}\BibitemShut {NoStop}%
\bibitem [{\citenamefont {Bondar}\ \emph {et~al.}(2016)\citenamefont {Bondar},
  \citenamefont {Cabrera}, \citenamefont {Campos}, \citenamefont {Mukamel},\
  and\ \citenamefont {Rabitz}}]{Bondar2014wigner}%
  \BibitemOpen
  \bibfield  {author} {\bibinfo {author} {\bibfnamefont {D.~I.}\ \bibnamefont
  {Bondar}}, \bibinfo {author} {\bibfnamefont {R.}~\bibnamefont {Cabrera}},
  \bibinfo {author} {\bibfnamefont {A.}~\bibnamefont {Campos}}, \bibinfo
  {author} {\bibfnamefont {S.}~\bibnamefont {Mukamel}}, \ and\ \bibinfo
  {author} {\bibfnamefont {H.~A.}\ \bibnamefont {Rabitz}},\ }\href@noop {}
  {\bibfield  {journal} {\bibinfo  {journal} {J. Phys. Chem. Lett.}\ }\textbf
  {\bibinfo {volume} {7}},\ \bibinfo {pages} {1632} (\bibinfo {year}
  {2016})}\BibitemShut {NoStop}%
\bibitem [{\citenamefont {Leggett}(1984)}]{leggett1984quantum}%
  \BibitemOpen
  \bibfield  {author} {\bibinfo {author} {\bibfnamefont {A.}~\bibnamefont
  {Leggett}},\ }\href@noop {} {\bibfield  {journal} {\bibinfo  {journal}
  {Physical Review B}\ }\textbf {\bibinfo {volume} {30}},\ \bibinfo {pages}
  {1208} (\bibinfo {year} {1984})}\BibitemShut {NoStop}%
\bibitem [{\citenamefont {Grabert}\ \emph {et~al.}(1984)\citenamefont
  {Grabert}, \citenamefont {Weiss},\ and\ \citenamefont
  {Hanggi}}]{grabert1984quantum}%
  \BibitemOpen
  \bibfield  {author} {\bibinfo {author} {\bibfnamefont {H.}~\bibnamefont
  {Grabert}}, \bibinfo {author} {\bibfnamefont {U.}~\bibnamefont {Weiss}}, \
  and\ \bibinfo {author} {\bibfnamefont {P.}~\bibnamefont {Hanggi}},\
  }\href@noop {} {\bibfield  {journal} {\bibinfo  {journal} {Phys. Rev. Lett.}\
  }\textbf {\bibinfo {volume} {52}},\ \bibinfo {pages} {2193} (\bibinfo {year}
  {1984})}\BibitemShut {NoStop}%
\bibitem [{\citenamefont {Pollak}(1986)}]{pollak1986transition}%
  \BibitemOpen
  \bibfield  {author} {\bibinfo {author} {\bibfnamefont {E.}~\bibnamefont
  {Pollak}},\ }\href@noop {} {\bibfield  {journal} {\bibinfo  {journal} {Phys.
  Rev. A}\ }\textbf {\bibinfo {volume} {33}},\ \bibinfo {pages} {4244}
  (\bibinfo {year} {1986})}\BibitemShut {NoStop}%
\bibitem [{\citenamefont {Leggett}(1996)}]{leggett1996effect}%
  \BibitemOpen
  \bibfield  {author} {\bibinfo {author} {\bibfnamefont {A.~J.}\ \bibnamefont
  {Leggett}},\ }in\ \href@noop {} {\emph {\bibinfo {booktitle} {Foundations Of
  Quantum Mechanics In The Light Of New Technology}}}\ (\bibinfo {organization}
  {World Scientific},\ \bibinfo {year} {1996})\ pp.\ \bibinfo {pages}
  {406--413}\BibitemShut {NoStop}%
\bibitem [{\citenamefont {Japha}\ and\ \citenamefont
  {Kurizki}(1996)}]{japha1996}%
  \BibitemOpen
  \bibfield  {author} {\bibinfo {author} {\bibfnamefont {Y.}~\bibnamefont
  {Japha}}\ and\ \bibinfo {author} {\bibfnamefont {G.}~\bibnamefont
  {Kurizki}},\ }\href {\doibase 10.1103/PhysRevLett.77.2909} {\bibfield
  {journal} {\bibinfo  {journal} {Phys. Rev. Lett.}\ }\textbf {\bibinfo
  {volume} {77}},\ \bibinfo {pages} {2909} (\bibinfo {year}
  {1996})}\BibitemShut {NoStop}%
\bibitem [{\citenamefont {Schaufler}\ \emph {et~al.}(1999)\citenamefont
  {Schaufler}, \citenamefont {Schleich},\ and\ \citenamefont
  {Yakovlev}}]{Schaufler1999keyhole}%
  \BibitemOpen
  \bibfield  {author} {\bibinfo {author} {\bibfnamefont {S.}~\bibnamefont
  {Schaufler}}, \bibinfo {author} {\bibfnamefont {W.~P.}\ \bibnamefont
  {Schleich}}, \ and\ \bibinfo {author} {\bibfnamefont {V.~P.}\ \bibnamefont
  {Yakovlev}},\ }\href {\doibase 10.1103/PhysRevLett.83.3162} {\bibfield
  {journal} {\bibinfo  {journal} {Phys. Rev. Lett.}\ }\textbf {\bibinfo
  {volume} {83}},\ \bibinfo {pages} {3162} (\bibinfo {year}
  {1999})}\BibitemShut {NoStop}%
\bibitem [{\citenamefont {Barone}\ \emph {et~al.}(2004)\citenamefont {Barone},
  \citenamefont {Kurizki},\ and\ \citenamefont {Kofman}}]{barone2004}%
  \BibitemOpen
  \bibfield  {author} {\bibinfo {author} {\bibfnamefont {A.}~\bibnamefont
  {Barone}}, \bibinfo {author} {\bibfnamefont {G.}~\bibnamefont {Kurizki}}, \
  and\ \bibinfo {author} {\bibfnamefont {A.~G.}\ \bibnamefont {Kofman}},\
  }\href {\doibase 10.1103/PhysRevLett.92.200403} {\bibfield  {journal}
  {\bibinfo  {journal} {Phys. Rev. Lett.}\ }\textbf {\bibinfo {volume} {92}},\
  \bibinfo {pages} {200403} (\bibinfo {year} {2004})}\BibitemShut {NoStop}%
\bibitem [{\citenamefont {Somma}\ \emph {et~al.}(2014)\citenamefont {Somma},
  \citenamefont {Reimann}, \citenamefont {Flytzanis}, \citenamefont
  {Elsaesser},\ and\ \citenamefont {Woerner}}]{somma2014high}%
  \BibitemOpen
  \bibfield  {author} {\bibinfo {author} {\bibfnamefont {C.}~\bibnamefont
  {Somma}}, \bibinfo {author} {\bibfnamefont {K.}~\bibnamefont {Reimann}},
  \bibinfo {author} {\bibfnamefont {C.}~\bibnamefont {Flytzanis}}, \bibinfo
  {author} {\bibfnamefont {T.}~\bibnamefont {Elsaesser}}, \ and\ \bibinfo
  {author} {\bibfnamefont {M.}~\bibnamefont {Woerner}},\ }\href@noop {}
  {\bibfield  {journal} {\bibinfo  {journal} {Phys. Rev. Lett.}\ }\textbf
  {\bibinfo {volume} {112}},\ \bibinfo {pages} {146602} (\bibinfo {year}
  {2014})}\BibitemShut {NoStop}%
\bibitem [{\citenamefont {Khamehchi}\ \emph {et~al.}(2017)\citenamefont
  {Khamehchi}, \citenamefont {Hossain}, \citenamefont {Mossman}, \citenamefont
  {Zhang}, \citenamefont {Busch}, \citenamefont {Forbes},\ and\ \citenamefont
  {Engels}}]{khamehchi2017negative}%
  \BibitemOpen
  \bibfield  {author} {\bibinfo {author} {\bibfnamefont {M.~A.}\ \bibnamefont
  {Khamehchi}}, \bibinfo {author} {\bibfnamefont {K.}~\bibnamefont {Hossain}},
  \bibinfo {author} {\bibfnamefont {M.~E.}\ \bibnamefont {Mossman}}, \bibinfo
  {author} {\bibfnamefont {Y.}~\bibnamefont {Zhang}}, \bibinfo {author}
  {\bibfnamefont {T.}~\bibnamefont {Busch}}, \bibinfo {author} {\bibfnamefont
  {M.~M.}\ \bibnamefont {Forbes}}, \ and\ \bibinfo {author} {\bibfnamefont
  {P.}~\bibnamefont {Engels}},\ }\href {\doibase
  10.1103/PhysRevLett.118.155301} {\bibfield  {journal} {\bibinfo  {journal}
  {Phys. Rev. Lett.}\ }\textbf {\bibinfo {volume} {118}},\ \bibinfo {pages}
  {155301} (\bibinfo {year} {2017})}\BibitemShut {NoStop}%
\bibitem [{\citenamefont {Larson}\ \emph {et~al.}(2005)\citenamefont {Larson},
  \citenamefont {Salo},\ and\ \citenamefont {Stenholm}}]{larson2005effective}%
  \BibitemOpen
  \bibfield  {author} {\bibinfo {author} {\bibfnamefont {J.}~\bibnamefont
  {Larson}}, \bibinfo {author} {\bibfnamefont {J.}~\bibnamefont {Salo}}, \ and\
  \bibinfo {author} {\bibfnamefont {S.}~\bibnamefont {Stenholm}},\ }\href
  {\doibase 10.1103/PhysRevA.72.013814} {\bibfield  {journal} {\bibinfo
  {journal} {Phys. Rev. A}\ }\textbf {\bibinfo {volume} {72}},\ \bibinfo
  {pages} {013814} (\bibinfo {year} {2005})}\BibitemShut {NoStop}%
\bibitem [{\citenamefont {Verlinde}(2011)}]{verlinde2011origin}%
  \BibitemOpen
  \bibfield  {author} {\bibinfo {author} {\bibfnamefont {E.}~\bibnamefont
  {Verlinde}},\ }\href@noop {} {\bibfield  {journal} {\bibinfo  {journal}
  {JHEP}\ }\textbf {\bibinfo {volume} {2011}},\ \bibinfo {pages} {1} (\bibinfo
  {year} {2011})}\BibitemShut {NoStop}%
\end{thebibliography}%

\end{document}